\documentclass[12pt]{article}
\newcommand{\beq}[3]{\begin{equation}  \label{#1#2#3}}
\newcommand{\eeq}{ \end{equation}}
\newcommand{\ba}{\begin{array}}
\newcommand{\ea}{\end{array}}

\newcommand{\remark}[1]{}
\let\LARGE=\Large
\let\Large=\large

\newcommand{\be}[3]{\begin{equation}  \label{#1#2#3}}     
\newcommand{\ee}{ \end{equation}}
\newcommand{\bea}{\begin{eqnarray}}
\newcommand{\eea}{\end{eqnarray}}

\newcommand{\ft}[2]{{\textstyle\frac{#1}{#2}}}

\def\beq{\begin{equation}}
\def\eeq{\end{equation}}
\def\beqa{\begin{eqnarray}}
\def\eeqa{\end{eqnarray}}

\begin{document}

\begin{titlepage}
\rightline{HU-EP-01/17}
\rightline{hep-th/0104156}

\vspace{15truemm}

\centerline{\bf \LARGE
Curved BPS domain wall solutions in
}
%
\vspace{3mm}
\centerline{\bf \LARGE
five-dimensional gauged supergravity }

\bigskip

\vspace{2truecm}

\centerline{\bf Gabriel Lopes Cardoso, Gianguido Dall'Agata
{\rm and} Dieter L\"ust}

\vspace{1truecm}

\centerline{\em  Institut f\"ur Physik, Humboldt University}
\centerline{\em Invalidenstra\ss{}e 110, 10115 Berlin, Germany}
\vspace{1truecm}
\centerline{\tt email:gcardoso,dallagat,luest@physik.hu-berlin.de}

\vspace{2truecm}


\begin{abstract}

We analyze the possibility of constructing supersymmetric curved
domain wall solutions in five-dimensional ${\cal N}=2$ gauged
supergravity, which are supported by non-constant scalar
fields belonging either to vector multiplets only or to vector and
hypermultiplets.
We show that the BPS equations for
the warp factor and for the vector scalars are modified by the presence
of a four-dimensional cosmological constant on the domain wall, in agreement
with earlier results by DeWolfe, Freedman, Gubser and Karch.
We also show that the cosmological constant on the domain wall is anti-de
Sitter like and that it constitutes an independent quantity, not related
to any of the objects appearing in the context of very special geometry.

\end{abstract}

\end{titlepage}


\newpage

\section{Introduction}

The idea \cite{Rubakov:1983bb} that our four-dimensional universe is
just a domain wall embedded in five-dimensional spacetime has
attracted a lot of attention recently
\cite{Lukas:1999yy}-\cite{Randall:1999vf}.
So far, the study of BPS domain wall solutions in
five-dimensional gauged supergravity has been restricted to the case
of flat four-dimensional domain walls
\cite{Lukas:1998tt}-\cite{CDKV}.
However, on general grounds, it is quite natural to ask whether there
also exist curved BPS domain walls in five-dimensional gauged
supergravity, i.e. four-dimensional curved domain walls with a
non-vanishing cosmological constant.  Such curved domain walls seem to
play an important role in the context of locally localized gravity
\cite{Karch:2000ct}-\cite{KR3} in spaces with infinite volume.

In four dimensions, curved
BPS domain wall solutions were recently constructed \cite{BCL} in
the context of ${\cal N}=2$ gauged supergravity coupled to vector
multiplets.
There it was shown that an anti-de Sitter cosmological constant on the
domain wall
is related to both the
imaginary part of the superpotential and to
a non-vanishing $U(1)$-connection.
Such a non-vanishing
connection is also known to give rise
to rotating four-dimensional BPS black hole solutions \cite{130,110}.
Since this $U(1)$-connection is expressed in terms of a set of harmonic
functions that also determine the behaviour of the scalar fields
belonging to the vector multiplets, the anti-de Sitter cosmological
constant on the
domain wall does not constitute an independent quantity.

On the other hand, in ${\cal N} = 2$ gauged supergravity in five
dimensions, there is no such $U(1)$-connection due to the reality
property of very special geometry and hence, a mechanism similar to the one
discussed above is not available.  Nevertheless,
as we will show in this paper, it may still
be possible to construct curved domain wall solutions which are supported
by non-constant scalar fields belonging either to vector multiplets only
or to
vector and hypermultiplets (we do not discuss the case of curved domain
wall solutions with non-constant hyper scalar fields only).
To achieve this, one has to give up the requirement that the
potential of the theory be written solely
in terms of a real superpotential $W$.
The resulting BPS equations for
the warp factor and for the vector scalars are then modified by the presence
of a four-dimensional cosmological constant on the domain wall.
These modified flow equations are in agreement with the earlier findings
of \cite{DeWolfe:2000cp}, which were established in the context of
non-supersymmetric gravitational theories in five dimensions.
We will also show that the cosmological constant on the domain wall
is then necessarily anti-de Sitter like and that it constitutes
an independent quantity, not related
to any of the objects appearing in
the context of very special geometry.  This is again in analogy with the
angular momentum carried by five-dimensional rotating BPS black hole
solutions in the theory of ${\cal N}=2$ ungauged supergravity coupled to
vector multiplets, where the angular momentum enters as an independent
quantity in the expression for the macroscopic Bekenstein-Hawking entropy
\cite{CS}.

In the case that all of the scalars are taken to be constant,
the resulting
solution describes
an ${\cal N}=2$ five-dimensional anti-de Sitter
vacuum with an (anti-)de Sitter foliation
along
the radial coordinate as in \cite{DeWolfe:2000cp}.

\section{Setup}

The five-dimensional ${\cal N}=2$ gauged supergravity theory that
we will consider in the following is the one constructed in
\cite{AnnaGianguido}, describing the general coupling of
$n_V$ abelian vector multiplets and of $n_H$
hypermultiplets to supergravity.
The scalar fields $\phi^x$ ($x=1, \dots, n_V$) of the vector multiplets
parametrize a very special manifold, whose sections $h^I(\phi)$
satisfy the constraint \cite{GST}
\bea
C_{IJK} h^I h^J h^K = 1 \;,
\label{v}
\eea
with real $C_{IJK}$ ($I = 0, \dots, n_V$).  The dual fields $h_I ( \phi)$
are defined by
\bea
h_I = C_{IJK} h^J h^K \;.
\label{dualh}
\eea
The hypermultiplet scalars $q^X$, on the other hand, parametrize a
quaternionic K\"{a}hler geometry determined by $4n_H$-beins
$f^{iA}_X(q^X)$, with the $SU(2)$ index $i=1,2$ and
the $Sp(2n_H)$ index $A=1,\ldots,2n_H$, raised and lowered by the
symplectic
metrics $C_{AB}$ and $\varepsilon_{ij}$.
We refer to \cite{AnnaGianguido} for more details.

We will be interested in the construction of BPS solutions with
spacetime metrics given by
\be100
ds^2 = {\rm e}^{2 U(r)} {\hat g}_{mn}
dx^m dx^n + {\rm e}^{-2 p U(r)} dr^2
\label{line}
\ee
with ${\hat g}_{mn} = {\hat g}_{mn} (x^m)$.  Here we denote spacetime
indices by $x^{\mu} = (x^m, r)$
and the
corresponding tangent space indices by $a = (0,1,2,3,5)$.  The
constant $p$ is introduced for later convenience.  We assume Lorentz
invariance in the four-dimensional subspace $a = (0,1,2,3)$.  The
metric ${\hat g}_{mn}$ is thus a four-dimensional constant curvature
metric,
\bea
{\hat R}_{mn} = - \ft{12}{l^2} \, {\hat g}_{mn} \;,
\label{rhat}
\eea
with the four-dimensional cosmological constant proportional to $l^{-2}$.
The case of (imaginary) real $l$ corresponds to a four-dimensional
(anti-) de Sitter spacetime.
We take the solutions to be uncharged, that is we set the
gauge fields to zero.  We allow for a non-trivial dependence of the scalar
fields on the coordinate $r$, and we write $\phi' = \partial \phi /\partial
r$, $q'= \partial q / \partial r$ as well as $U' = \partial U / \partial r$.

In the absence of gauge fields, the supersymmetry
transformation laws for the gravitini $\psi_{\mu i}$, for
the gaugini $\lambda^x_i$
and for the hyperini
$\zeta^A$ read
\cite{AnnaGianguido,CDKV}
\begin{eqnarray}
\delta
\psi _{\mu i}  & = & { D}_\mu (\omega )\epsilon
_i+ \, \ft{i}{\sqrt{6}}\; g\; \gamma _\mu \; P_{ij} \; \epsilon^j,
\label{gravi} \\
 \delta
\lambda _i^{x}&=&- \ft{i}{2}(\not\!{\partial}
\phi^{x})\epsilon _i
+g \; P_{ij}^{x}\; \epsilon^j \,, \label{gaugi}\\
 \delta
\zeta^A &=& - \ft{i}{2} f_{iX}^A(\not\!{\partial}
  q^{X})\epsilon ^i + g  \; {\cal N}_i^A\; \epsilon^i\,,
 \label{transfos1}
\end{eqnarray}
where (as for all triplets $P_{ij}=i P^s(\sigma ^s)_{ij}$)
 \begin{eqnarray}
   P^s &\equiv & h^I(\phi) P_{I}^{s}(q)\,, \qquad P_{x}^s \equiv
-\sqrt{\ft{3}{2}} \partial_{x} P^s= h^I_{x}P_{I}^{s}\,, \nonumber\\
{\cal N}_i^A & \equiv&\ft{\sqrt{6}}{4} f_{iX}^AK^X  \,,\qquad K^X\equiv
h^I(\phi
)K_I^X(q)\,.
 \label{defcN}
\end{eqnarray}

The scalar potential of such a theory is given by the squares of the
shifts of the supersymmetry transformations
 \begin{equation} \label{potential}
{\cal V} =- 4 P^s P^s +2 P_{x}^s P_{y}^s
g^{xy} + 2 {\cal N}_{iA}{\cal N}^{iA} \,
 \end{equation}
and depends only on the triplet of prepotentials $P^s$ and their
derivatives.  By
decomposing the triplet $P^s$ into its norm and phases,
\begin{equation}
  P^s=\sqrt{\ft32} W Q^s\,,\qquad Q^s Q^s=1\,,
 \label{PWQ}
\end{equation}
one obtains
\bea
W = h^I ( \phi) \; P_I   \;\;\;,\;\;
P_I  \equiv  \sqrt{\ft{2}{3}} \, P^s_I (q) \,Q^s \;
\label{pi}
\eea
as well as \cite{CDKV}
\begin{equation}
 {\cal V}= - 6 \, W^2+\ft{9}{2} \, g^{\Lambda\Sigma}\partial_\Lambda
 W\partial_\Sigma W\, + \ft{9}{2}\,  W^{2}
 (\partial _x Q^s)(\partial ^x Q^s) \;. \label{pot}
\end{equation}
Here $g_{\Lambda \Sigma}$ denotes the metric of the complete scalar
manifold, which is positive definite,
involving the scalars of both the vector and the hypermultiplets.
The derivatives acting on  the $SU(2)$ phases $Q^s$ are just computed
with respect to the scalars of the vector multiplets.
As noted in \cite{CDKV}, when $\partial^x Q^s= 0$, i.e. when the
phases only depend on the quaternions (or when they are constant), then
the potential
goes into a form for which gravitational stability
is guaranteed \cite{Town}.

We can now compute the integrability conditions coming from the vanishing
of the gravitini variation (\ref{gravi}).
As already stated above, we only allow for a non-trivial dependence
of the scalar fields on the coordinate $r$.
This results in the following integrability conditions (for $p=0$):
\bea
R_{mn} &=& \left[4 g^2 W^2 -
\ft13
\phi^{x\prime}  \phi^{y\prime} g_{xy} (\phi)  -\ft16  q^{X\prime}
q^{Y\prime} g_{XY}(q) - \ft14 K^X K^Y g_{XY}(q)\right] \, {\rm e}^{2U} \,
\hat g_{mn},
\\
R_{rr} &=& 4 g^2 W^2 -
\ft43
\phi^{x\prime}  \phi^{y\prime} g_{xy} (\phi)  -\ft23  q^{X\prime}
q^{Y\prime} g_{XY}(q) -  K^X K^Y g_{XY}(q) \;,
\eea
where we also made use of the vanishing of the transformations
(\ref{gaugi}) and (\ref{transfos1}).

For a metric of the form (\ref{100}) subject to (\ref{rhat}) this yields
\bea
3 U^{''} +\ft{12}{l^2}\, {\rm e}^{-2U} &=& - \phi^{x\prime}\phi^{y\prime}
g_{xy} -  \ft12 q^{X\prime} q^{Y\prime} g_{XY} - \ft92 \, g^2
g^{XY} \, \partial_X W \partial_Y W \;,
\label{eins1}\\
(U^\prime)^2 - \ft{4}{l^2}\, {\rm e}^{-2U} &=&  g^2 W^2 \,, \label{eins2}
\eea
where we used that $ g_{XY} \, K^X K^Y =
6  g^{XY} \, \partial_X W \partial_Y W$.
For supersymmetric configurations,
these equations are equivalent to the Einstein equations.  It follows
that the BPS equations we will obtain below must be compatible
with (\ref{eins1}) and (\ref{eins2}).  {From} (\ref{eins2}) we obtain
that
\bea
U' = \pm \gamma(r) \, g W \;,
\label{modwarp}
\eea
where
\bea
\gamma(r)=\sqrt{1+\frac{4 {\rm e}^{-2U}}{ l^2 g^2W^2}}\;.
\label{gamma}
\eea
The first order differential equation (\ref{modwarp})
for the warp factor ${\rm e}^U$
was also derived in \cite{DeWolfe:2000cp} in the context of
non-supersymmetric five-dimensional gravity theories.

\section{Flat BPS domain walls}

The case of flat BPS domain walls without hypermultiplets has
already been discussed extensively in the literature
\cite{BC1}-\cite{280}.  Flat BPS domain walls
with hypermultiplets have been discussed
in \cite{BG,B,CDKV}.
Here we briefly review some of the features of these solutions,
and we present an extension of these results to the case of non-trivial
hypermultiplets.

Inserting the spacetime line element (\ref{line}) into the gravitini
variation
yields
\bea
\delta \psi_{m i} &=& {\hat {\cal D}}_{m} \epsilon_i
+ \ft{1}{2} {\rm e}^{p U} \partial_r U \gamma_m \gamma_5 \epsilon_i
+ \ft{i}{2}  \, g W Q_{ij}
 \, \gamma_{m} \, \epsilon^j \;, \label{gravm} \\
\delta \psi_{r i} &=& \partial_r \epsilon_i - q^{X \prime} \;
\omega_{X\,i}{}^j \epsilon_j
+ \ft{i}{2}  \, g W {\rm e}^{- p U} Q_{ij}
 \, \gamma_{5} \, \epsilon^j \;,
\eea
where ${\hat {\cal D}}_{m}$ denotes the covariant derivative with
respect
to the metric $\hat g_{mn}$.

First consider the variation
$\delta \psi_{m i} =0$.  Since the four-dimensional domain wall is flat
($\gamma =1$),
we
set
\bea
{\hat {\cal D}}_{m} \epsilon_i = 0\, .\label{flatsusy}
\eea
We then obtain from $\delta\psi_{mi}=0$ that
\bea
 \ft{1}{2} {\rm e}^{p U} \partial_r U  \gamma_5
\epsilon_i = - \ft{i}{2} g W Q_{ij} \epsilon^j \;.
\label{grav1}
\eea
The consistency of
(\ref{grav1})
implies that
\bea
 {\rm e}^{2p U} \Big( \partial_r U \Big)^2 = g^2 W^2 \;,
\eea
and hence
\bea
{\rm e}^{p U} \partial_r U  = \pm g W \;,
\label{warp}
\eea
which is in accordance with (\ref{modwarp}).
Inserting (\ref{warp}) into (\ref{grav1}) yields the projector condition
\bea
i \gamma_5 \epsilon_i = \pm Q_{ij} \epsilon^j \;.
\label{projector}
\eea
Since this is the only condition we impose on the supersymmetry
parameters, we conclude that the resulting flat domain wall solutions
preserve $1/2$ of ${\cal N}=2$ supersymmetry.

Next, let us consider the vanishing of the variation of the other
fermion fields.
Using the projector equation (\ref{projector}), it follows
from (\ref{gaugi}) and (\ref{transfos1}) that the
scalar fields $ \phi^{\Lambda} = \{\phi^x, q^X\}$ have to satisfy
\cite{CDKV}
\begin{equation}
  \phi^{\Lambda '}=\mp 3 g \, {\rm e}^{-pU}
\, g^{\Lambda \Sigma }\partial _\Sigma
W\,.
 \label{qprimeall}
\end{equation}
We note that (\ref{qprimeall}) and (\ref{warp}) imply
(\ref{eins1}).

Consistency of the above equations also gives the following constraint
on the phases
$Q^s$ \cite{CDKV}
\bea
\partial_{x} Q^s = 0.
\eea
Thus, it follows from (\ref{pi}) that $W = h^I( \phi) P_I (q)$.
Using the chain rule $\partial_r h_{I} = \partial_x h_I
\phi^{x \prime}$ as well as (\ref{warp}),
one then obtains from (\ref{qprimeall}) that
\bea
{\rm e}^{p U} \partial_r h_I (\phi)  + 2  {\rm e}^{p U} (\partial_r U)
h_I (\phi)  = \pm 2 g P_I (q) \;.
\label{gaug2}
\eea
We note that (\ref{gaug2}) implies (\ref{warp}).  This follows simply by
contracting (\ref{gaug2}) with $h^I$ and using that $h^I h_I =1$ as well
as its consequence $h^I \partial_r h_I =0$.

Next, we
introduce the rescaled fields
\bea
Y^I = {\rm e}^U h^I \;\;\;,\;\;\;
Y_I = {\rm e}^{2U} h_I
\;,
\label{yx}
\eea
It then follows
from (\ref{v}) and (\ref{dualh}) that
\bea
{\rm e}^{3U} = C_{IJK} Y^I Y^J Y^K \;\;\;,\;\;\;
Y_I = C_{IJK} Y^J Y^K \;.
\label{u}
\eea
Setting
\bea
p = 2
\eea
for convenience \cite{BG}, we then obtain from (\ref{gaug2}) that
\bea
\partial_r Y_I = \pm 2 g P_I(q)  \;,
\eea
which is solved by
\bea
Y_I =  c_I \pm 2 g \int^r   dr' \, P_I(q)  \;,
\eea
where the real $c_I$ denote integration constants.
A solution to this equation in closed form can only be given if
one determines the behaviour of the quaternions $q^X$.
It does not appear to be possible to determine the behaviour
of $P_I (q)$ in a way that is independent from the vector multiplet scalars.

When only vector multiplets are present, and in the case
of Abelian gaugings, the superpotential $W$
reduces to $W = h^I \alpha_I$, where $\alpha_I$ are the
Fayet-Iliopoulos terms, which are constant.
Then it follows that \cite{BG}
\bea
Y_I =  c_I \pm 2 g \alpha_I r
 \;.
\label{attractor}
\eea
These equations  are called attractor equations.

\section{Curved BPS domain walls}

Now we take $\hat g_{mn}$ to be a constant curvature metric satisfying
(\ref{rhat}).
Then (\ref{flatsusy})
gets modified, and a natural ansatz describing this modification is given by
\bea
{\hat {\cal D}}_{m} \epsilon_i = f(r) ~\hat{e}_m{}^a
\gamma_a \gamma_5 \epsilon_i +\frac{i}{2}h(r)Q_{ij}\hat{e}_m{}^a\gamma_a
\epsilon^j
\,,
\label{antidesusynew}
\eea
where $f(r)$ and $h(r)$ denote real functions.  Equation (\ref{antidesusynew})
gives rise to the following integrability condition:
\bea
f^2(r) - \frac{1}{4}h^2(r)=\frac{1}{l^2}\, .\label{intnew}
\eea
As we shall see momentarily,
the simple choices $f^2 = l^{-2}$, $h=0$ and $f=0$, $h^2 = - 4
l^{-2}$ for the case of real and imaginary $l$, respectively, do not lead
to a supersymmetric solution.

We set $p=0$ in the following.
Inserting (\ref{antidesusynew}) into
the supersymmetry condition $\delta \psi_{m i} =0$ now yields
\bea
\Big({\rm e}^{-U} f(r) + \ft{1}{2}
\partial_r U \Big)
\gamma_5
\epsilon_i = - \ft{i}{2} \Big( {\rm e}^{-U} h(r)+gW \Big)
\, Q_{ij} \epsilon^j \, .
\label{grav2}
\eea
The consistency of (\ref{grav2}) implies that
\bea
{\rm e}^{-U} f(r) + \ft{1}{2}
\partial_r U
= \pm \ft{1}{2} \Big( {\rm e}^{-U} h(r) + gW \Big) \;.
\label{modcons}
\eea
Then, combining (\ref{modcons}) with
(\ref{intnew}) and using (\ref{modwarp})
yields
\bea
f(r)= \mp \ft{1}{2} {\rm e}^U \gamma \, g W \;\;\;,\;\;\;
h(r)= - {\rm e}^{U} g W
\, ,
\label{fr}
\eea
with $\gamma$ given by (\ref{gamma}).
We note that when inserting (\ref{fr}) into (\ref{grav2}) both sides of
the latter equation vanish, so that (\ref{grav2}) does not result in a
restriction of the amount of supersymmetry preserved by the solution.

In the case of constant scalar fields,
the superpotential $W$ is constant and equation
(\ref{modwarp}) can be solved explicitly.  For $\gamma \neq 0$ and
$|l| \neq \infty$ the solution reads
\bea
{\rm e}^U= \frac{2}{l g W } \sinh \Big(\pm g W\, (r - r_0) \Big)\,
\eea
for real $l$, whereas for imaginary $l$ it is given by
\bea
{\rm e}^U= \frac{2}{|l| g W } \cosh \Big(\pm g W\, (r - r_0) \Big)\, .
\eea
As shown in \cite{DeWolfe:2000cp}, this describes
a four-dimensional (anti-)de Sitter foliation of
the ${\cal N}=2$
supersymmetric $AdS_5$ vacuum.  It can be checked that (\ref{eins1})
is satisfied by this solution.
The case $\gamma =0$, on the other hand,
does not lead to a solution of (\ref{eins1}).

Let us now turn to the case of non-constant scalar
fields.  First consider the case when $\partial_x Q^s = 0$.
Below we will show that, in order to
construct non-trivial solutions to the gaugino
variation equation $\delta\lambda_i^x=0$
preserving 1/2 of ${\cal N}=2 $ supersymmetry,
one has to impose the
projector condition (\ref{projector}) on the supersymmetry parameters.
Inspection of (\ref{grav2}) then shows that $f = h = 0$, so that $l^{-1}=0$
and $\gamma =1$, which gives rise to a flat domain wall solution.

Thus, in order to obtain curved
BPS domain wall solutions with either non-constant vector scalar fields
or non-constant vector and hyper
scalar fields\footnote{It may be possible to construct curved domain wall
solutions
with only non-constant hyper scalars turned on \cite{BC}.},
one has to allow for a dependence of the $Q^s$ on some of
the vector multiplet scalars\footnote{In ${\cal N}=2$ gauged
supergravity coupled to vector multiplets only, it may be possible to
achieve $\partial_x Q^s \neq 0$ by a non-abelian gauging.}.
We will shortly see that this implies that the projector condition
(\ref{projector})
has to get modified, as follows:
\bea
i \gamma_5 \epsilon_i = A(r) \; {Q_i}^j \epsilon_j + B(r) \; {M_i}^j
\epsilon_j \;.
\label{procmod}
\eea
The $SU(2)$-valued matrix $Q = i Q^s \sigma^s$ contains
the phases of the prepotentials $P_{ij}$
(as in (\ref{PWQ})), which may now
depend on both the vector and the hyper scalars, and
$M = i M^s \sigma^s$
denotes an $SU(2)$-valued matrix satisfying $M_i\,^j M_j\,^k =- \delta_i\,^k$
(i.e. $M^s M^s =1$).
Without loss of generality, we take $Q$ and $M$ to be orthogonal in $SU(2)$
space, so that $Q^s M^s = 0$.
The consistency of (\ref{procmod}) then yields that
\bea
A^2(r) + B^2(r) &=& 1 \;.\label{AB}
\eea
Since (\ref{procmod}) is the only condition we will impose
on the supersymmetry parameters, the resulting curved domain wall solutions
will preserve 1/2 of ${\cal N}=2$ supersymmetry.

Inserting (\ref{procmod}) into the gaugini variation equation
$\delta \lambda_i^x =0$ yields (with $p=0$)
 \begin{equation}
\Big( A(r) \; {Q_i}^j \epsilon_j + B(r) \; {M_i}^j \epsilon_j \Big)
\; \phi^{x \prime} =  \sqrt{6} g
 \, g^{xy}
\partial_{y} P_i\,^j \epsilon_j\,.
\label{Gauginotransf}
\end{equation}
Then, using the decomposition of $P^s$ given in (\ref{PWQ}) yields
\begin{equation}
\Big(A(r) \; Q^s +B(r) \; M^s \Big) \phi^{x \prime} =
3g \, g^{xy} (Q^s \partial_{y} W + W \partial_{y} Q^s).
\end{equation}
Since $Q^s \partial_{x}  Q^s=0$, the two pieces on the right hand side of
this equation are
orthogonal to each other, and hence it follows that
\begin{equation}
A(r) \phi^{x \,\prime} =  3 g\, g^{xy} \partial_{y} W\,
\label{dWvector}
\end{equation}
as well as
\begin{equation}
B(r) M^s \phi^{x \prime} = 3 \,g W \; g^{xy} \partial_{y} Q^s\,.
\label{dQr0}
\end{equation}
Inspection of (\ref{dQr0}) shows that if the $Q^s$ do not depend on any
of the vector scalar fields, then either all the vector scalar fields are
constant, or $B M^s =0$.  In the latter case,
the projector condition
(\ref{procmod}) reduces to (\ref{projector}), thus yielding flat domain wall
solutions.  Hence we take
$\partial_{\tilde x} Q^s \neq 0$
in the following.  We denote
the subset of vector scalar
fields, on which the $Q^s$ depend, by $\phi^{\tilde x}$.
The remaining vector scalars will be
denoted by $\phi^{\hat x}$.
In order to be able to solve
(\ref{dWvector}) and (\ref{dQr0}), we take the metric $g_{xy}$ to be
factorisable as $g_{xy} = (g_{\hat x \hat y}, g_{\tilde x \tilde y})$.
Then, we find from (\ref{dWvector}) and (\ref{dQr0}) that
\bea
\phi^{\hat x} = {\rm constant} \;\;,\;\; \partial_{\hat x} W = 0 \;.
\label{fixsca}
\eea
The scalar fields $\phi^{\tilde x}$, on the other hand, are non-constant.

Squaring (\ref{dQr0}) and using $M^s M^s =1$ yields
\bea
B^2(r) =
9 g^2W^2 \,  \frac{ g^{\tilde x \tilde y}
(\partial_{\tilde x}
Q^s) (\partial_{{\tilde y}}
Q^s) }{g_{{\tilde x}{\tilde y}} \,
\phi^{{\tilde x}
\prime} \phi^{{\tilde y} \prime} } \;.
\label{b21}
\eea
Since the right hand side of (\ref{b21})
is positive
definite due to the reality of the $Q^s$ and the positivity of the
metric $g_{\tilde x \tilde y}$,
we conclude that $B(r)$ is a real function.

On the other hand, we obtain from (\ref{dQr0}) that
\begin{equation}
    \label{Mr}
M^s = \frac{3 g W}{B} \frac{\phi^{{\tilde x} \prime} \partial_{\tilde x}
Q^s}{g_{{\tilde x} {\tilde y}} \, \phi^{{\tilde x}
\prime} \phi^{{\tilde y} \prime} } \;.
\end{equation}
Squaring this and using again that $M^s M^s =1 $ yields
\bea
B^2(r) = 9 g^2W^2 \,  \frac{ \left(\phi^{{\tilde x} \prime} \partial_{\tilde x}
Q^s ) (\phi^{{\tilde y} \prime} \partial_{{\tilde y}}
Q^s\right)}{ (
g_{{\tilde x} {\tilde y}} \, \phi^{{\tilde x}
\prime} \phi^{{\tilde y} \prime})^2
} \;.
\label{b2}
\eea
Equating (\ref{b2}) with (\ref{b21}) then yields
\bea
(\phi^{{\tilde x} \prime} \partial_{\tilde x}
Q^s ) (\phi^{{\tilde y} \prime} \partial_{{\tilde y}}
Q^s )= (g_{{\tilde x}{\tilde y}} \, \phi^{{\tilde x}
\prime} \phi^{{\tilde y} \prime} ) \,  (g^{\tilde z \tilde w}
(\partial_{\tilde z}
Q^s) (\partial_{{\tilde w}}
Q^s) ) \;.
\label{bb}
\eea
We note that in the case that $Q^s$ only depends on one scalar field
$\phi^{\tilde x}$, then the relation (\ref{bb}) is automatically
satisfied.

We now determine $A(r)$ by using the integrability
condition resulting from the gravitini equation $\delta \psi_{mi} =0$.
Inserting (\ref{procmod}) into (\ref{gravm}) yields
\bea
\hat{\cal D}_m \epsilon_i = \ft{i}{2} \left( U^\prime \, A\, + g W
\right) \gamma_m {Q_i}^j \epsilon_j + \ft{i}{2} U^\prime B
\gamma_m  {M_i}^j
\epsilon_j \;,
\eea
whose integrability gives
\bea
\left( U^\prime A + g W\right)^2 + B^2 (U^\prime)^2 = - 4
l^{-2} {\rm e}^{-2U} \;.
\label{relab}
\eea
Inserting (\ref{AB}) as well as (\ref{modwarp})
into (\ref{relab}) then gives
\bea
A(r) = \mp \gamma (r) \;.
\label{ag}
\eea
Since the left hand side of (\ref{relab}) is positive definite,
we conclude that $l$ has to be purely imaginary.  Thus we conclude that
the presence of non-constant vector scalars excludes the possibility of
having BPS de Sitter like domain walls.

Inserting (\ref{ag}) into (\ref{dWvector}) yields
\begin{equation}
\phi^{{\tilde x} \,\prime} = \mp 3 g \, \gamma^{-1} \, g^{\tilde x \tilde y}
\partial_{\tilde y} W\;.
\label{flowmod}
\end{equation}
Comparison of (\ref{qprimeall}) with (\ref{flowmod}) shows that,
in the presence of an anti-de Sitter
cosmological constant on the wall,
the flow equation for the vector scalar fields $\phi^{\tilde x}$
gets modified by a factor $\gamma^{-1}$.  This is in agreement with
a similar finding \cite{DeWolfe:2000cp} in the context of non-supersymmetric
gravity theories in five dimensions.

We note that in the case of constant hyper scalars, (\ref{flowmod}) and
(\ref{modwarp}) imply (\ref{eins1}).

Using the chain rule $\partial_r h_I = \partial_x h_I \phi^{x'} =
\partial_{\tilde x} h_I \phi^{{\tilde x}'} $ as well as (\ref{modwarp}),
one obtains from (\ref{flowmod}) that
\bea
\partial_r h_I (\phi)  + 2 \gamma^{-2}  (\partial_r U)
h_I (\phi)  = \pm 2 g \gamma^{-1} P_I  \;.
\label{hp}
\eea
We note that (\ref{hp}) implies (\ref{modwarp}).  This follows simply by
contracting (\ref{hp}) with $h^I$ and using that $h^I h_I =1$ as well
as $h^I \partial_r h_I =0$.

Introducing the rescaled fields
\bea
Y^I = {\rm e}^{\int^r dr' \gamma^{-2}(r') (\partial_{r'} U)
}\; h^I \;\;\;,\;\;\;
Y_I = {\rm e}^{2\int^r  dr' \gamma^{-2}(r') (\partial_{r'} U) } \;h_I
\label{rescyg}
\eea
yields
\bea
\partial_r Y_I = \pm 2 g \gamma^{-1}  \,
{\rm e}^{2\int^r dr' \gamma^{-2}(r') (\partial_{r'} U)} \; P_I \;.
\eea
Under the coordinate change $dr \rightarrow
{\rm e}^{-2 \int^r dr' \gamma^{-2}(r') (\partial_{r'} U)
}\; dr$ this goes into
\bea
\partial_r Y_I = \pm 2 g \gamma^{-1}  \,P_I \;,
\eea
and hence
\bea
Y_I = c_I \pm 2 g \int^r dr'  \gamma^{-1}(r')  \,
 P_I \;,
\eea
where the real $c_I$ denote integration constants.
An equation for the warp factor analogous to (\ref{u}) can be obtained by
inserting (\ref{rescyg}) into (\ref{v}).  An explicit solution for the warp
factor may then be derived by expanding in powers of $l^{-1}$.

We also obtain from (\ref{b21}), (\ref{ag}), (\ref{flowmod})
and (\ref{AB}) that
\bea
 g^{\tilde x \tilde y}
(\partial_{\tilde x}
Q^s) (\partial_{{\tilde y}}
Q^s ) = g^{\tilde x \tilde y}
(\partial_{\tilde x}
W) (\partial_{{\tilde y}}
W)
\, \frac{(1 - \gamma^2)}{ \gamma^2 \,
W^2} \;.
\label{QQ}
\eea
Inserting (\ref{QQ}) into the expression for the potential (\ref{pot})
then yields
\begin{equation}
 {\cal V}= - 6 \, W^2+\ft{9}{2} \,\gamma^{-2} \;
g^{\tilde x \tilde y}\partial_{\tilde x}
 W\partial_{\tilde y} W\, +
\ft{9}{2} \, g^{X Y}\partial_X
 W\partial_Y W\,
\;,
\label{potgubs}
\end{equation}
where we also used (\ref{fixsca}).  The expression (\ref{potgubs}) for the
potential is in agreement with the one given in \cite{DeWolfe:2000cp}.

%

Finally, let us briefly comment on
the vanishing of the hyperini variation, $\delta
\zeta^A =0$.  The insertion of (\ref{procmod}) into
$\delta \zeta^A =0$ results in a more complicated matrix equation than
(\ref{qprimeall}), with the
metric $g^{XY}$ in (\ref{qprimeall})
replaced by a more complicated object.  We leave the analysis
of the resulting matrix equation for the future.

\medskip

To summarize, we have addressed the issue of the construction
of curved BPS domain wall solutions in five dimensions.  We have seen
that in the case when such domain wall solutions are supported
either by non-constant vector scalars or
by both non-constant vector and hyper scalars, we have to require
$\partial_{\tilde x} Q^s \neq 0$.
We have also seen that only anti-de Sitter domain walls are allowed
by this construction.
The resulting first order BPS equations for the warp factor and for the
vector scalars ((\ref{modwarp}) and (\ref{flowmod}), respectively)
are modified by the factor $\gamma$ which depends on the
four-dimensional cosmological constant on the wall.  These equations agree
with those obtained previously in \cite{DeWolfe:2000cp} in the context
of non-supersymmetric five-dimensional gravity theories.
We also note the presence of the
additional condition (\ref{bb}) on the solution, as well as those arising
from the flow equations for the hyper scalars which we haven't analyzed
in this paper.

In the presence
of hypermultiplets, it has been shown \cite{CDKV} that a gauged supergravity
theory can
possess critical points with ultraviolet and infrared directions.
Using the AdS/CFT correspondence, this leads to the possibility of the
construction of domain wall solutions dual to regular renormalization
group flows.  It would be very interesting to see if it
is possible to construct curved BPS domain wall solutions
interpolating between such vacua and to understand the meaning of the
domain wall cosmological constant in the dual field theory flow.
As an example one could consider deforming the two-parameter
solution of \cite{CDKV} describing the ${\cal N}=2$ embedding of the
UV-IR solution of \cite{FGPW}.

\bigskip

{\bf Acknowledgements}

We would like to thank K. Behrndt,
P. Breitenlohner, B. de Wit and D. Maison for valuable discussions.
We are very much indepted to A. Karch, L. Randall and A. Strominger
for pointing out an error in the previous version of this paper.
 This work
is supported by the DFG and by the European Commission RTN Programme
HPRN-CT-2000-00131.





\providecommand{\href}[2]{#2}\begingroup\raggedright

\endgroup

\end{document}